# Explainable Machine Learning and Deep Learning Models for Predicting TAS2R-Bitter Molecule Interactions

Francesco Ferri[1,2†], Marco Cannariato[1†], Lorenzo Pallante[1], Eric A. Zizzi[1], and Marco A. Deriu[1*]

[1] Politecnico di Torino, PolitoBIOMedLab, Department of Mechanical and Aerospace Engineering, Torino, 10129, Italy

[2] Leibniz Institute for Food Systems Biology at the Technical University of Munich, 85354 Freising, Germany

[†] These authors contributed equally to this work

* marco.deriu@polito.it

## Abstract

This work aims to develop explainable models to predict the interactions between bitter molecules and TAS2Rs via traditional machine-learning and deep-learning methods starting from experimentally validated data. Bitterness is one of the five basic taste modalities that can be perceived by humans and other mammals. It is mediated by a family of G protein-coupled receptors (GPCRs), namely taste receptor type 2 (TAS2R) or bitter taste receptors. Furthermore, TAS2Rs participate in numerous functions beyond the gustatory system and have implications for various diseases due to their expression in various extra-oral tissues. For this reason, predicting the specific ligand-TAS2Rs interactions can be useful not only in the field of taste perception but also in the broader context of drug design. Considering that in-vitro screening of potential TAS2R ligands is expensive and time-consuming, machine learning (ML) and deep learning (DL) emerged as powerful tools to assist in the selection of ligands and targets for experimental studies and enhance our understanding of bitter receptor roles. In this context, ML and DL models developed in this work are both characterized by high performance and easy applicability. Furthermore, they can be synergistically integrated to enhance model explainability and facilitate the interpretation of results. Hence, the presented models promote a comprehensive understanding of the molecular characteristics of bitter compounds and the design of novel bitterants tailored to target specific TAS2Rs of interest.

## Introduction

Taste perception is a crucial determinant of food intake and consumption patterns (Glendinning, 1994), with substantial consequences for human nutrition and health (Shahbandi et al., 2018). Several pathological conditions can impair the sense of taste and lead to alterations in quality of life and body weight regulation (Risso et al., 2020). The molecular mechanisms of bitter taste perception involve a subfamily of 25 G protein-coupled receptors (GPCRs), called type 2 taste receptors (TAS2Rs) (Pallante et al., 2021). These proteins are also expressed in various extraoral tissues, where they may perform additional physiological functions, such as modulating inflammatory response, controlling upper respiratory immunity and others (Behrens & Lang, 2022; Behrens & Meyerhof, 2011). TAS2Rs exhibit a remarkable diversity and specificity in their ligand recognition and activation. Some TAS2Rs are defined as *promiscuous* as they can bind to multiple and structurally distinct bitter compounds, while others are highly *selective* and respond to only a few known ligands (Di Pizio & Niv, 2015). Conversely, some bitter compounds can activate several TAS2Rs, while others are specific for individual TAS2Rs. The chemical nature of bitter compounds is extremely heterogeneous and encompasses peptides, saponins, alkaloids, polyphenols, and salts (Di Pizio et al., 2019). In this context, understanding the molecular interactions and predicting the specific association between bitter molecules and relative TAS2Rs could impact several fields of applications. For example, this line of research can: (i) help in the adherence to therapies based on bitter drugs (Mennella et al., 2013); (ii) improve the understanding regarding side effects or affect undiscovered biological pathways since TAS2Rs are present in extraoral tissues (Shaik et al., 2016); (iii) assist the design of alternative bitterants to improve the palatability of some healthy foods with high level of bitterness (Sun-Waterhouse & Wadhwa, 2013). However, the current methods to identify the TAS2R targets for a given compound rely on laborious and costly in-vitro assays and, for this reason, the available information in the literature regarding protein-ligand interactions for bitter receptors is limited. Still, most of the data has been collected into a single database, called BitterDB (Dagan-Wiener et al., 2019).

To overcome the above-mentioned limitations, several computational approaches to predict the bitter taste or the interaction between bitter compounds and TAS2Rs have been proposed (Malavolta et al., 2022). Computational methods offer several advantages, such as speed, low cost, and scalability, as well as the possibility of improvement as more experimental data become available. Currently, many machine learning (ML) models have been proposed to accurately classify a compound as bitter or non-bitter, e.g. (Zheng et al., 2018), BitterIntense (Margulis et al., 2021), BitterCNN (Bo et al., 2022), and VirtuousSweetBitter (Maroni et al., 2022). Moreover BitterX (Huang et al., 2016), BitterSweet (Tuwani et al., 2019), and BitterMatch (Margulis et al., 2022) can also predict the specific TAS2R target (F. Ferri et al., 2024). BitterX uses a Support Vector Machine (SVM) trained on a reduced and balanced dataset, whereas BitterMatch employs Gradient Boosting (GB) on Decision Trees (DTs) trained on data extracted from BitterDB (Dagan-Wiener et al., 2019). On the other hand, although the BitterSweet webserver offers information on bitterant-TAS2R associations within its application, there is a lack of detail in both the original publication and the online materials regarding the specific development of this predictive model for association. While the above-mentioned algorithms rely on traditional ML, in recent years, Deep Learning (DL) models based on Neural Networks (NNs) have often offered increased performance. In particular, Graph Neural Networks (GNNs) are becoming widely used in computational chemistry since they are particularly suitable for the representation of molecular data (Guo et al., 2023). In general, the common drawback of DL and many ML algorithms is their "black-box" nature which results in a complex explanation of the model's predictions and feature importance. Therefore, several methods have been developed for both traditional ML algorithms and GNNs (Lundberg et al., 2019; Pope et al., 2019; Ying et al., 2019). Among methods to explain traditional machine-learning methods, SHAP is a widely known model-agnostic method that can help in the understanding of how the features influence the prediction by elucidating toward which class the prediction is pushed (Lundberg & Lee, 2017). On the other hand, tree-based algorithms can immediately output the feature importance but do not provide any further information. Specifically, Graph Neural Networks (GNNs) can directly represent molecular structures by encoding atoms and bonds as nodes and edges, respectively. This enables a more straightforward explanation of the model's predictions on the molecular structures. For instance, both model-agnostic GNNExplainer (Ying et al., 2019) and Grad-CAM (Selvaraju et al., 2020) methods unveil the influence of the single nodes and edges in the class decision, granting more visually impactful explanations.

In this work, we present two complementary models capable of predicting interactions between bitter compounds and TAS2Rs using traditional machine learning (ML) methods and graph neural networks (GNNs). These models could assist in identifying bitter compounds that selectively target specific bitter taste receptors and shed light on the molecular features underlying these interactions. Ultimately, this research could advance our understanding of the molecular features behind bitter taste perception and facilitate the development of tailored compounds designed to target specific TAS2Rs.

## Material and Methods

### Dataset acquisition and preprocessing

The dataset comprises 338 bitter molecules with known positive and negative associations with 22 human bitter taste receptors (the remaining three receptors, namely TAS2R45, TAS2R48, and TAS2R60, are orphan and do not have any known agonists). Positive associations correspond to molecule-receptor couples that are known to interact, while negative interactions refer to molecules that do not bind to the associated receptor. Positive interactions are labelled as class "1" and negative as class "0". Uniquely known and in-vitro verified interactions were considered, resulting in a total of 3964 paired associations. More in detail, 301 molecules (3204 known association with TAS2Rs) were taken from BitterMatch's dataset (Margulis et al., 2022), which is in turn derived from the BitterDB dataset (Dagan-Wiener et al., 2019). The remaining 37 molecules (760 known association with TAS2Rs) were derived from recent literature (Behrens et al., 2018; Cui et al., 2021; Delompré et al., 2022; Jaggupilli et al., 2019; Karolkowski et al., 2023; Lang et al., 2020; Morini et al., 2021; Nouri et al., 2019; Soares et al., 2018). The resultant dataset exhibits a remarkable imbalance, with the number of binding instances approximately five times greater than the number of non-binding instances. In summary, the final dataset therefore considered 22 out of the 25 TAS2R receptors and consisted of 3964 pairs of bitterants-TAS2Rs known associations. The problem of predicting ligand-receptor association was reduced to a binary classification by defining the entries as bitterant-TAS2R pairs and the association as the only label. The resulting dataset is represented as a matrix where the number of rows equals the number of known associations. Each row contains information about the structure of the bitter compound, the associated receptor, and the target label, indicating the positive or negative association. Molecules are encoded as Canonical SMILES obtained either by BitterDB or PubChem (Kim et al., 2023). Receptors are instead defined using the one-hot encoding strategy.

### Interaction prediction using a traditional machine-learning (TML) approach

Molecules SMILES were standardized following the ChEMBL structure pipeline (Bento et al., 2020), as done in previous literature (Maroni et al., 2022; Pallante et al., 2022). Fingerprints and physicochemical descriptors were used as features for the bitter molecules. Morgan fingerprints (number of bits = 1024, radius = 2) were computed using RDKit python package starting from the previously sanitized SMILES. Molecular descriptors were calculated with the Mordred Python library (Moriwaki et al., 2018). All descriptors having more than 90% correlation with other descriptors were removed. All non-binary data was normalized with Min-Max normalization (Patro & Sahu, 2015) (see also Supplementary Information for further details on the normalization procedure).

Several traditional machine learning algorithms, namely Gaussian Naive Bayes (GaussianNB), Logistic Regression (LR), K-Neighbors, Support Vector Machines (SVM), Random Forest (RF), and Gradient Boosting on Decision Trees (GB on DTs), were first compared. The model demonstrating the highest performance (GB on DTs) was subsequently chosen as the optimal model for further analyses. This algorithm was also used in similar models (Margulis et al., 2022) and demonstrated remarkable resilience towards overfitting and biases, which are essential when dealing with imbalanced data (Natekin & Knoll, 2013). CatBoost (Dorogush et al., 2018), a high-performance open-source library for gradient boosting on decision trees, was employed. CatBoost uses a combination of ordered boosting, random permutations, and gradient-based optimization to achieve high performance on any dataset. The comprehensive workflow for the TML approach is shown in Figure 1.

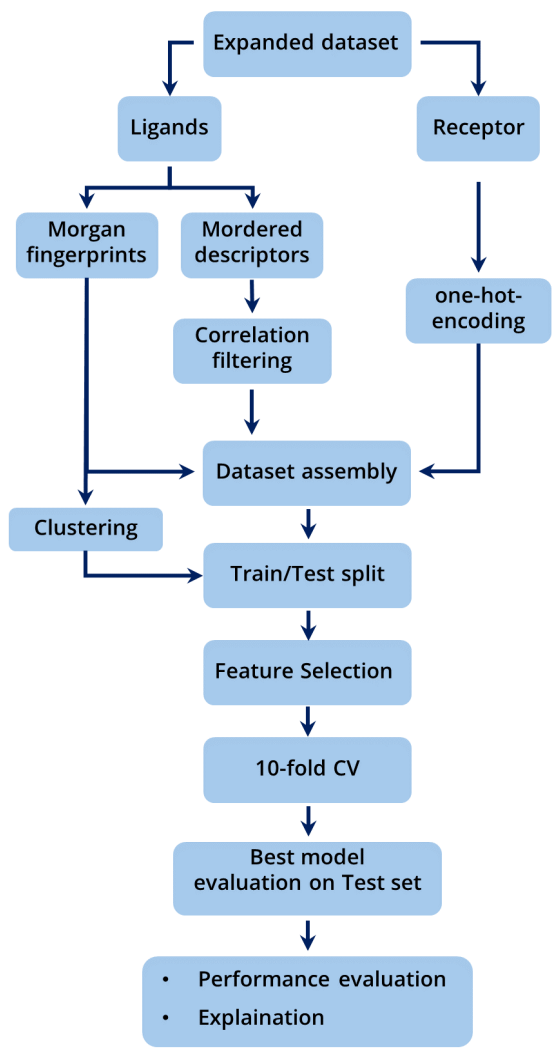

*Figure 1. Traditional Machine-Learning (TML) workflow.*

We performed data clustering before splitting the dataset into training and test sets to ensure a good representation of the chemical space in both sets. We used agglomerative clustering (complete linkage algorithm) to divide data into $n$ clusters with similar variances. To define the optimal $n$ number of clusters the Silhouette score analysis over different $n$ values was carried out. The parameter chosen as the distance ($d$) to divide the chemical space into clusters with Agglomerative clustering was the Tanimoto distance (Rogers & Tanimoto, 1960), computed from the precomputed Morgan fingerprints. The entries from each cluster were then split into training and test tests (80:20) by stratifying over the class labels.

The total number of ligand-based features was 2824. Dimensionality reduction of the dataset is, therefore, necessary to keep only the most informative features and simplify the model explanation. Feature selection was performed using two methods, i.e. (i) the "*noisy*" feature selection approach and (ii) Sequential Feature Selection (SFS). The noisy feature selection method is based on previous literature (Akhiat et al., 2021) and adopts an iterative selection technique by incorporating a random column into the dataset. Termed "noisy", this column is populated with pseudo-random numbers ranging from 0 to 1. Following the training phase of the tree-based classifier, the Gini importance for each feature was computed. Features with lower importance than the noisy feature were then systematically excluded from the dataset until only features more informative than the noisy column remained. The SFS is a greedy algorithm, that selects a subset of features in a forward or backward direction. The selection or removal of a feature depends on the cross-validation (CV) score of an estimator trained on the current feature subset. Forward-SFS begins with no features and gradually adds the best feature that maximizes the CV score until it reaches the desired number of features. Similarly, the Backwards-SFS starts with all features and progressively removes the worst feature that minimizes the CV score (F. J. Ferri et al., 1994). Scikit-learn was employed to perform backward-SFS considering the average precision as the 5-fold CV score, starting from the 150 most important features according to the CatBoostClassifier tree-based importance metric. The final number of features was decided a posteriori.

Finally, the classification task was then carried out by training the CatBoostClassifier on the training set, using a 10-fold CV, and evaluating the model's performance on the test set.

**Interaction Prediction using Graph Convolutional Neural Networks**

Graph Convolutional Networks (GCNs) extend the neural network models to work on arbitrarily structured graphs using graph convolutions and spectral filters (Kipf & Welling, 2017). A GCN is an approach for semi-supervised learning on graph-structured data. It is based on an efficient variant of convolutional neural networks which operate directly on graphs. In particular, spatial-based methods define graph convolutions by using the node's spatial relations similarly to the convolutional operator of a CNN on an image. The proposed model is built on PyTorch and PyTorch Geometric (Fey & Lenssen, 2019; Paszke et al., 2019).

Referring to the Graph as $G = (V, E)$, $V$ is the set of nodes and $E$ is the set of edges. Suppose $v_i$ is a node in $V$ and $e_{ij} = (v_i, v_j)$ is an edge that originates from $v_i$ and terminates at $v_j$. The graph is therefore defined by $n$ nodes, $m$ edges, and $d, b,$ and $c$ as dimensions of the node, hidden node, and edge feature vectors, respectively (Wu et al., 2021). The set of nodes that are adjacent to a node $v$ is denoted by $N(v) = \{u \in V | (v, u) \in E\}$. The graph has an adjacency matrix $A$ of size $n \times m$, such that $A_{ij} = 1$ if and only if $e_{ij}$ belongs to $E$, and $A_{ij} = 0$ otherwise. The graph also has node features $X$, which is a $n \times d$ matrix of node feature vectors, each of which has $d$ dimensions and corresponds to a node $v$, i.e., $x_v$. Additionally, the graph may have an edge feature matrix $X^e$, of dimension $m \times c$, composed by edge feature vectors of $c$ dimensions and represents an edge $(v, u)$, i.e. $x^e_{v,u}$. A molecule is defined as an undirected graph in which the adjacency matrix is always symmetric.

Molecules, stored as standardized SMILES, were converted into molecular graphs using NetworkX (Hagberg et al., 2008). The node and edge features were selected according to previous literature (He et al., 2024; Lim et al., 2019; Nguyen et al., 2021) and are listed in Table S1. As in the TML approach, the dataset was clustered based on the Tanimoto similarity before splitting each cluster into training and test tests (80:20) by stratifying over the class labels. The developed model takes as input a batch of graphs, each with node features, edge features, and the one-hot encoded receptor, and predicts the positive or negative association with the specified receptor. The overall workflow for the GCN approach is shown in Figure 2.

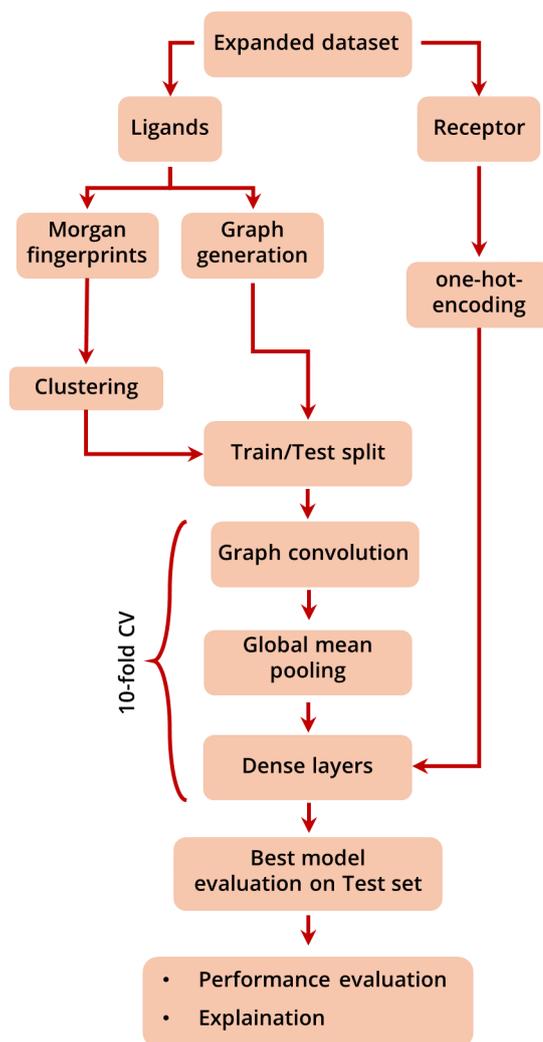

*Figure 2. GCN framework workflow.*

The GCN model consists of four main components: the graph convolutional layers, the batch normalization layers, the dropout layers, and the fully connected layers. The graph convolutional layers use the GATv2Conv module, which implements the Graph Attention Network (GAT) variant proposed by previous literature (Veličković et al., 2018). The GAT module computes the node embeddings by applying a self-attention mechanism over the node features and the edge features. The model uses two graph convolutional layers, with 32 and 8 output channels, respectively. Batch normalization layers are placed after both convolutional layers. Graph embeddings are then mapped using four fully connected layers, with 32, 16, 8, and 4 output units, respectively. The model uses two dropout layers, one with a probability of 0.1 and one with a probability of 0.2. The first dropout layer is applied to the input of the fully connected layers, and the second dropout layer is applied to the output of the last fully connected layer. The first fully connected layer takes as input the concatenation of the graph embeddings, obtained by applying global mean pooling to the node embeddings, and the receptor features. The last fully connected layer is followed by an output layer, which uses a linear transformation to produce two outputs. The model makes use of the ReLU activation function for the hidden units and of a sigmoid activation function for the node embeddings. The model returns the output of the output layer, which can be interpreted as the probability of each class. The architecture of the GCN model is schematically represented in Figure 3.

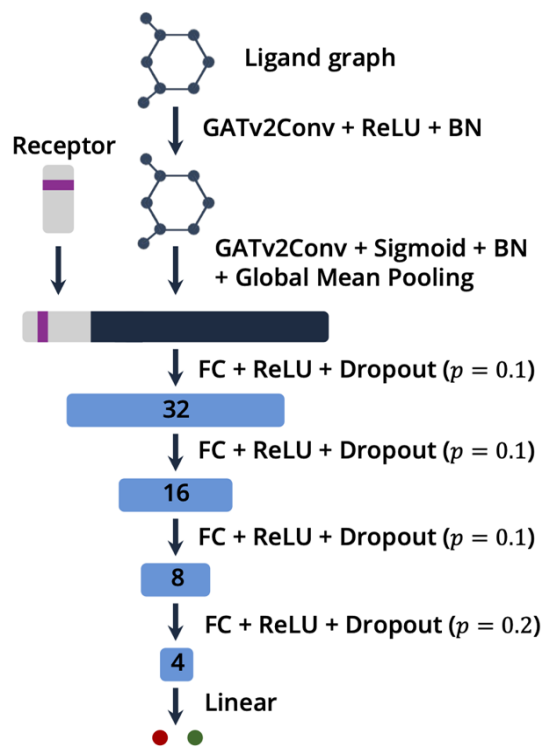

*Figure 3. Schematic representation of the GCN model architecture.*

**Explainability**

Regarding the TML, CatBoost evaluates the relevance of each input feature by using the individual importance values. These values reflect the average variation in the prediction caused by the modification of the feature value. Moreover, we employed SHAP (SHapley Additive exPlanations) to underscore the specific feature importance (Lundberg & Lee, 2017). SHAP values are a way to interpret the output of any machine learning model, based on the principles of game theory. SHAP values exhibit consistency, indicating that features deemed unequivocally more important will consistently yield higher SHAP values. The employed SHAP library uses a Tree-based model as the local explanation method for trees (Lundberg et al., 2019), allowing the calculation of optimal local explanations.

The explainability of the GCN model was achieved through GNNExplainer (Ying et al., 2019) and Grad-CAM (Selvaraju et al., 2020) methods. GNNExplainer is a model-agnostic approach for providing interpretable explanations for predictions of any GNN-based model on any graph-based machine learning task (Ying et al., 2019), allowing for single-instance explanations. At the same time, the unfaithfulness of the explanation was evaluated using the graph explanation faithfulness (GEF) score (Agarwal et al., 2022), calculated as:

$$GEF(y, \hat{y}) = 1 - e^{-KL(y||\hat{y})}$$

where $y$ and $\hat{y}$ refer to the output probability vector obtained from the original graph and masked subgraph, respectively, and $KL$ is the Kullback-Leibler divergence score. Therefore, the GEF score ranges between 0 and 1, with values near 0 indicating excellent prediction faithfulness and values close to 1 indicating very poor faithfulness. Typically, faithfulness scores higher than 0.5 are considered indicative of untrustworthy explanations.

On the other hand, Grad-CAM, originally developed to identify salient regions in image classification problems (Selvaraju et al., 2020), considers the gradient of the output with respect to the last convolutional layer. In this work, a generalization to graphs in the form of Unsigned Grad-CAM (UGrad-CAM) was employed to identify positive and negative contributions from each node (Pope et al., 2019). The implementation of both Grad-CAM and UGrad-CAM was adapted from publicly available code repositories (https://github.com/ndey96/GCNN-Explainability).

# Results

**Interaction prediction using a traditional machine-learning (TML) approach**

Initially, we compared traditional machine learning algorithms, including Gaussian Naive Bayes (GaussianNB), Logistic Regression (LR), K-Neighbors, Support Vector Machines (SVM), Random Forest (RF), and Gradient Boosting on Decision Trees (GB on DTs). GB on DT achieved better performances in terms of ROC (Receiver Operating Characteristics) AUC (Area Under the Curve) (see also Figure S1) and was therefore selected as the best model for the traditional machine learning approach. The CatBoostClassifier hyperparameters were tuned on the training set and are detailed in Table S2. Then, two feature selection methods, namely the "noisy" and the Backward-sequential feature selection (SFS) methods, were compared to select only the most informative ligand-based features, as also detailed in the Materials and Methods section. 28 ligand features were selected using the "noisy" method, whereas 17 through the SFS approach (Figure S2). The two feature selection methods achieved similar performance in terms of ROC and Precision-Recall (PR) AUC on the test set (Figure S3). It is interesting to note that both methods ultimately selected only Mordered descriptors, while no ligand fingerprints were retained in the final feature set. Given the higher reproducibility and the lower number of selected features, the SFS method was selected as the most suitable for the feature selection task. Therefore, from now on, the GB on DT using the SFS method be referred to as the *TML model*. The relative ROC and PR curves on the test set, reported in Figure 4, were characterized by AUC values of 0.92 and 0.75, respectively.

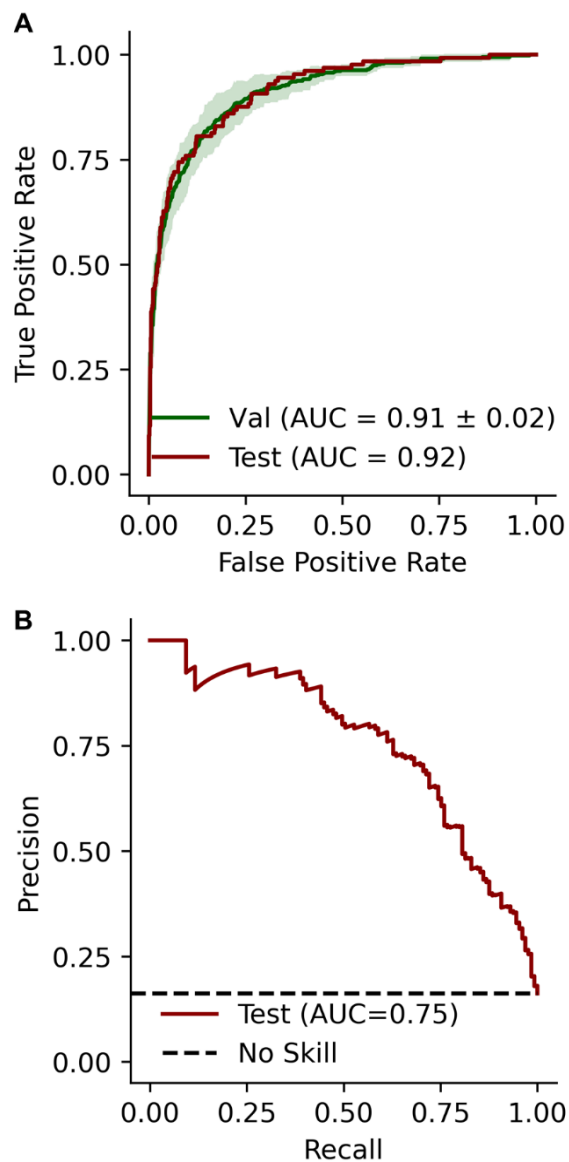

*Figure 4. (A) TML's ROC curves for the validation (green, mean and standard deviation during the 10-fold CV), and test (red) sets. (B) PR curve on the test set.*

The tree-based importance of the employed features was evaluated (Figure 5). The importance of the Mordered descriptors is in general the highest, with the only exception of the association with TAS2R14 and TAS2R46, which are the first and third most important features according to the model. Notably, these TAS2Rs are also the two most promiscuous receptors.

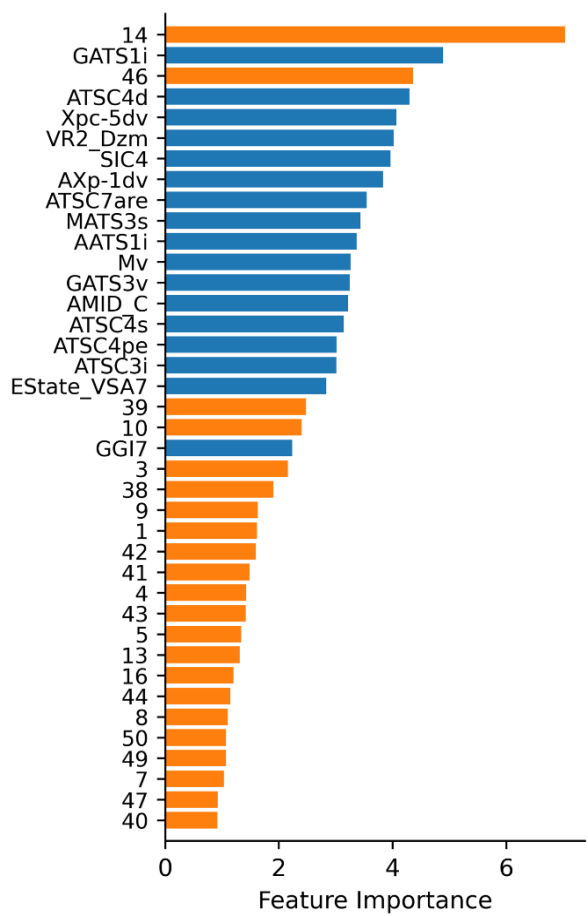

*Figure 5. Tree-based feature importance of the TML model. The 22 receptor association features are displayed in orange, while the 17 ligand descriptors selected by the SFS method are shown in blue.*

The feature importance of the TML model was also assessed through the SHAP method (Figure 6). Comparing the SHAP values for the one-hot encoded receptor features (Figure 6B) to the number of agonists of each TAS2R (Figure 6A), it could be observed that associations with the most promiscuous receptors tend to bias the prediction toward class 1, whereas connections with more selective receptors lean towards the class 0. Moreover, SHAP was employed to evaluate how specific features influence individual predictions. For example, the explanation for the strychnine-TAS2R46 pair (positive association) and the strychnine-TAS2R1 pair (negative association) are shown in Figure 6C,D. It is noteworthy that the association feature with the respective receptor typically holds a notable influence on the prediction outcome. However, in certain instances, such as observed with the strychnine-TAS2R1 pair, other features may also remarkably impact the final predicted association.

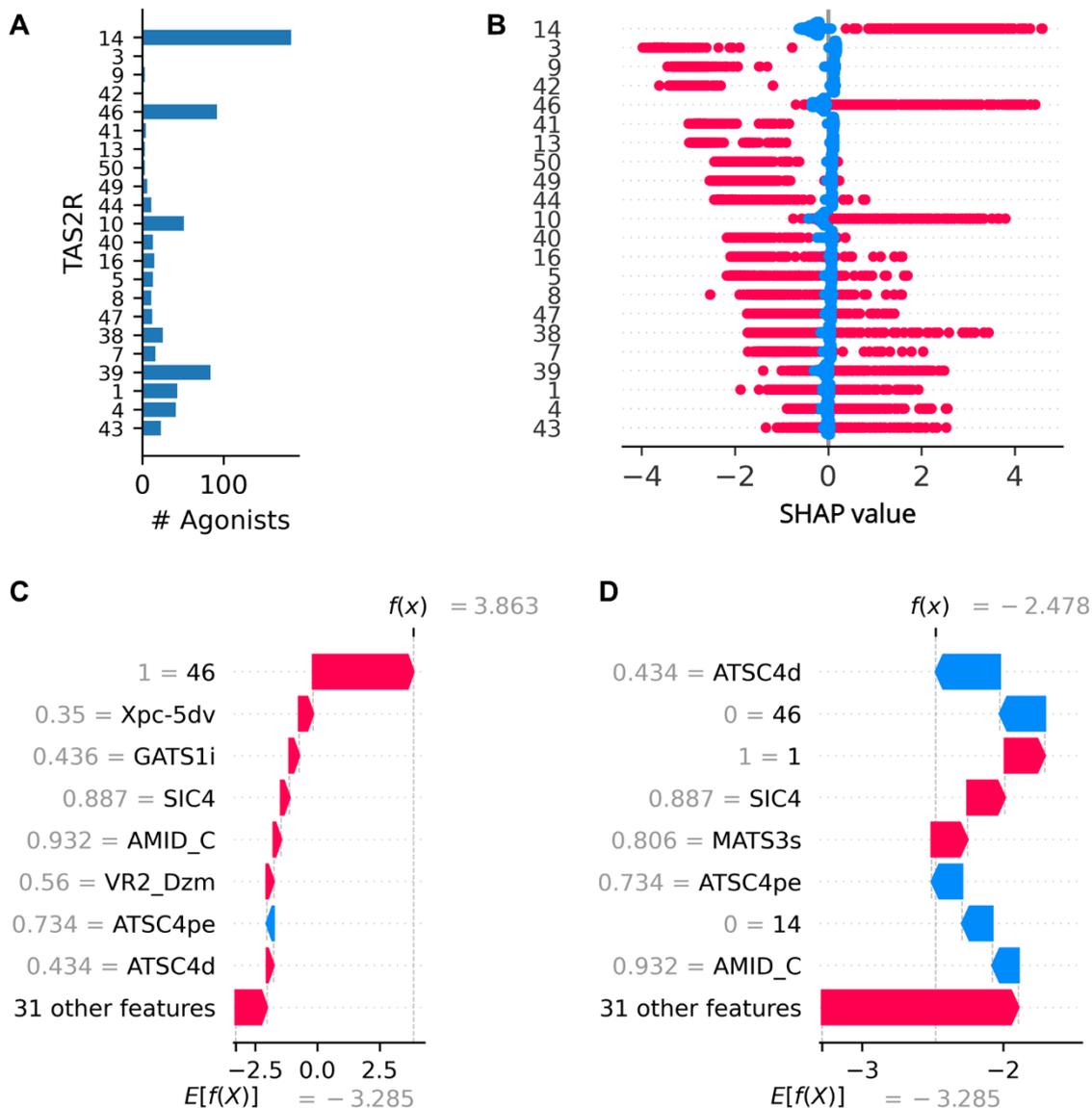

*Figure 6. (A) Number of agonists per TAS2R and (B) relative SHAP values for receptor associations. (C, D) SHAP waterfall for strychnine-TAS2R46 (positive association) (C) and strychnine-TAS2R1 (negative association) (D) pairs. E[f(X)] is the average raw prediction (log odds) across the dataset and f(x) is the prediction for the specific pair (log odds). Blue bars correspond to negative contributions to the prediction, while red bars to positive contributions. Features and their value for the specific pair are reported on the vertical axis.*

### Interaction prediction using Graph Neural Network approach

Similarly to the TML model, the ROC and PR curves for the GCN classifier were generated by conducting a 10-fold CV and selecting the model with the highest performance in the best fold to evaluate it on the Test Set (Figure 7). The GCN model exhibited a ROC AUC of 0.88, demonstrating performance on the Test set comparable to that observed during the CV. Regarding the PR curve, the model was characterized by an AUC of 0.67.

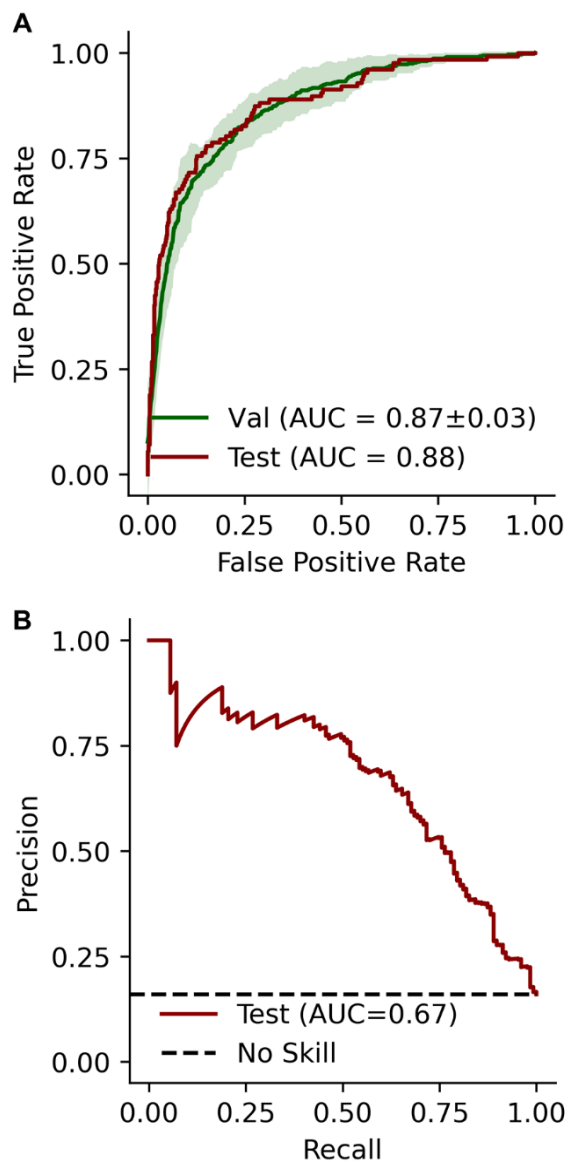

*Figure 7. (A) GCN's ROC curves for the validation (green, mean and standard deviation during the cross-validation), and test (red) sets. (B) PR curve on the test set.*

We then employed the GNNExplainer (Ying et al., 2019) and Grad-CAM (Selvaraju et al., 2020) methods to develop single-instance explanations for the GCN model predictions. Considering the case of the strychnine-TAS2R46 association (class 1) as an example, the GNNExplainer made it possible to inspect the importance of node features for the prediction (Figure 8A) as well as of each bond within the molecule (Figure 8B). Notably, the three most important node features include the atom's partial charge and partition coefficient, which can be related to the hydrophilicity of the molecule. Moreover, the bonds formed by a tertiary amine, which was experimentally found to be involved in an interaction with TAS2R46, are among the most relevant for the prediction. On the other hand, UGrad-CAM analysis was used to inspect the contribution of each node in the prediction toward the two classes. Such information, for the strychnine-TAS2R46 pair, is shown as a heatmap in Figure 8C. Interestingly, the region involving the aforementioned tertiary amine holds a great contribution toward class 1. On the other hand, the aromatic ring of strychnine represents the main contribution towards class 0. This explainability approach can be also useful to inspect the possible effect of structural alterations of the molecule on the output prediction. For instance, removing two carbon atoms from strychnine near the above-mentioned tertiary amine changes the pattern of importance values (Figure 8D). Such modification results in a contribution toward class 0 for this amine, which becomes secondary, together with a reduced influence for other regions of the molecule. As a result, the output probability of the model was reduced.

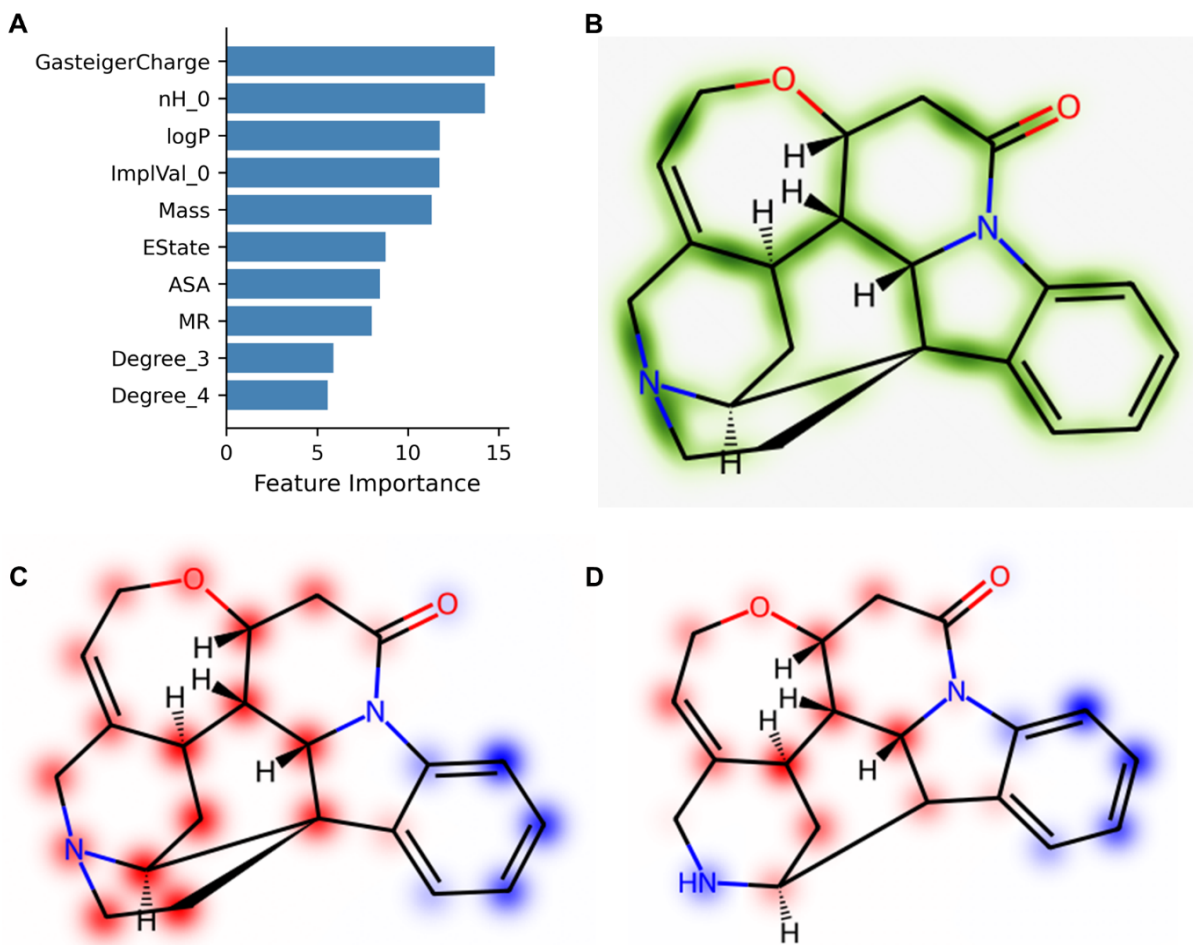

*Figure 8. GCN model explainability for the strychnine-TAS2R46 pair (positive association) using GNNExplainer (A, B) and UGrad-CAM (C, D). (A) Node feature importances for the 10 most important features for the prediction and (B) heatmap of edge importance, where the darkness of the green is proportional to the importance of the edge. (C) UGrad-CAM heatmap where red nodes correspond to contribution towards class 1 and blue towards class 0. (D) UGrad-CAM heatmap of a modified strychnine molecule paired to the TAS2R46 receptor, where red nodes correspond to contribution towards the predicted class (1), and blue towards the opposite class (0).*

Finally, the performance of the TML and GCN models on the test set was compared (Table 1). Despite GCN achieving higher recall and F2 scores on class 1 compared to the TML model, the TML model is characterized, on average, by higher values of the considered metrics. In particular, the TML precision is remarkably higher than the GCN precision for class 1. Therefore, regarding the under-represented class, TML seemed to prioritize the precision while GCN the recall.

*Table 1. Comparison of TML and GCN performance metrics on the Test set. Values in bold represent the highest of the metrics compared.*

|  | ROC AUC | PR AUC | Class | Precision | Recall | F1 | F2 |
|---|---|---|---|---|---|---|---|
| **TML** | **0.92** | **0.75** | 0 | 0.93 | **0.97** | **0.95** | **0.96** |
|  |  |  | 1 | **0.78** | 0.60 | **0.68** | 0.63 |
| **GCN** | 0.88 | 0.67 | 0 | **0.94** | 0.92 | 0.93 | 0.93 |
|  |  |  | 1 | 0.62 | **0.67** | 0.64 | **0.66** |

## Discussion

The present work aims to develop a machine-learning-based model to predict which TAS2R receptor is targeted by a specific bitter molecule based on its molecular structure. To achieve this goal, two complementary approaches using traditional machine learning methods and graph-based neural networks have been developed.

The dataset consists of 343 bitter molecules for which the positive (binders) or negative (non-binders) association with the human TAS2R bitter taste receptors is experimentally known. The resulting dataset is therefore composed of 3970 bitterant-TAS2R pairs with a relative label corresponding to a positive (class 1) or negative (class 0) interaction. A major obstacle to the present work is the paucity, diversity and unbalancing of the available data on TAS2R-ligand interactions. The dataset contains only bitter molecules, limiting the applicability of the models only to this specific taste. Regarding the applicability domain (AD) of the models, we evaluated the reliability of the model predictions based on the similarity between the tested compounds and the chemicals used during the training phase. We used an average-similarity approach already employed in previous literature in the taste prediction field (Zheng et al., 2018, 2019) and in our previous works (Maroni et al., 2022; Pallante et al., 2022) (see also Supplementary Information and Figure S4). The AD check is performed every time before running the model to assess the reliability of the prediction and the output of the AD control is given to the user. Future experimental studies elucidating the interactions between other bitter compounds or other chemicals with TAS2Rs would considerably enhance the performance of the models and broaden their chemical applicability domain. Moreover, the models could be improved by adding features related to the three-dimensional structure of the bitter taste receptors that are known to drive the ligand binding and recognition, such as the volume of the binding pocket, the Solvent Accessible Surface Area (Area), the radius of gyration, and many others highlighted in previous literature (Margulis et al., 2022). Indeed, the accurate experimental or *in silico* determination of the molecular structures of bitter receptors, akin to GPCRs in general, remains a complex and ongoing challenge in this area of research.

After comparing several traditional machine learning algorithms, Gradient Boosting on Decision Trees (GB on DTs) was chosen as the best architecture for the present problem (see also Figure S1). Interestingly, the RandomForest achieved similar performance to the GB on DTs, confirming the soundness of tree-based algorithms for solving this specific prediction problem. It is noteworthy that even the most recent comparable work in the literature, such as BitterMatch, relies on a tree-based method (XGBoost) (Margulis et al., 2022). To reduce the number of ligand-based features, the GB on DT was then tested with two feature selection methods: (i) the "noisy" feature selection method, which preserves only features with importance greater than a random-generated feature; and (i) the Backward Sequential Feature Selection (SFS) method, which evaluate the performance of the model by iteratively removing a feature from the dataset. The performance with the two feature selection methods showed similar results both on ROC and PR curves (see also Figure S3). For this reason, despite its higher computational cost, the Backward-SFS was preferred to the "noisy" feature selection method since it is completely reproducible and since it allowed to select only 17 features, compared to the 28 selected by the "noisy" method.

A tree-based algorithm also facilitated the assessment of feature importance, as shown in Figure 5. The most important feature is the association with receptor 14, noticeably the most promiscuous receptor with the highest number of known agonists, followed by receptor 46, which is the second most numerous in terms of known binders. As clearly shown in the bar plot, the more selective is the receptor, the less the importance value of their association. On the other hand, regarding the ligand-based descriptors, the SFS method selected 17 features, comprising only Mordered molecular descriptors and no ligand fingerprints. More in detail, the most informative features are: GATS1i (geary coefficient of lag 1 weighted by ionization potential), ATSC4d (centered moreau-broto autocorrelation of lag 4 weighted by sigma electrons), Xpc-5dv (5-ordered Chi path-cluster weighted by valence electrons), VR2_Dzm (normalized Randic-like eigenvector-based index from Barysz matrix weighted by mass), SIC4 (5-ordered structural information content), AXp-1dv (1-ordered averaged Chi path weighted by valence electrons), ATSC7are (centered moreau-broto autocorrelation of lag 7 weighted by allred-rocow EN), MATS3s (moran coefficient of lag 3 weighted by intrinsic state), AATS1i (averaged moreau-broto autocorrelation of lag 1 weighted by ionization potential), Mv (mean of constitutional weighted by vdw volume), GATS3v (geary coefficient of lag 3 weighted by vdw volume), AMID_C (averaged molecular ID on C atoms), ATSC4s (centered moreau-broto autocorrelation of lag 4 weighted by intrinsic state), ATSC4pe (centered moreau-broto autocorrelation of lag 4 weighted by pauling EN), ATSC3i (centered moreau-broto autocorrelation of lag 3 weighted by ionization potential), EState_VSA7 (EState VSA Descriptor 7 ( 1.81 <= x < 2.05)), and GGI7 (7-ordered raw topological charge). The comprehensive compilation of employed Mordred descriptors can be accessed at https://mordred-descriptor.github.io/documentation/master/descriptors.html. The descriptors with higher occurrence fall into the categories of autocorrelation and topological descriptors. Autocorrelation descriptors (GATS1i, ATSC4d, ATSC7are, ATSC4s, ATSC4pe, ATSC3i) capture spatial relationships between atoms or properties within a molecule, while topological (Xpc-5dv, MATS3s, GGI7) descriptors focus on the molecular graph's structure. Despite the substantial reduction in the number of features, comprehending the chemical and physical properties of tastants based on the 17 most important features remains challenging. To improve the model's explainability, future studies should prioritize the utilization of simpler descriptors or the development of specific methodologies to intuitively relate the molecular descriptors with the relative structural features or functional groups. In this context, the explainability obtained for the GCN model is considerably easier to understand and to directly correlate molecular features to the overall prediction.

Moreover, SHAP values highlighted not only the importance of the single features on the final prediction but also their specific influence on the final prediction. Firstly, Figure 6A,B shows that pairs involving more promiscuous receptors (like TAS2R14, 46, and 39) pull the prediction toward the positive class. On the other hand, being associated with a selective receptor favours the negative class. SHAP feature importance also allows to obtain a local interpretation of the model, analysing a specific bitterant-TAS2R association. For example, Figure 6C shows how the feature's value influences the prediction of the strychnine-TAS2R46 pair (known association: positive – class 1, prediction: class 1), while Figure 6D for strychnine-TAS2R1 (known association: unknown, prediction: class 0). In the case of the strychnine-TAS2R46 pair, the feature associated with TAS2R46 exerts a substantial influence, driving the prediction towards the positive class. This finding aligns with expectations, as TAS2R46 is recognised as one of the most promiscuous bitter taste receptors, and the dataset consequently incorporates numerous instances of interacting compounds. Conversely, for the strychnine-TAS2R1 interaction, the most influential feature is a ligand-based descriptor (ATSC4d), suggesting that the molecular structure of the ligand is the primary factor driving the prediction in this instance. It is important to underline that due to the dataset's imbalance, with approximately five times as many negative samples as positive ones, the overall prediction tends to skew towards the negative class. Specifically, lower negative SHAP values exert a stronger influence on pushing predictions towards the negative class, while lower positive SHAP values primarily serve to counterbalance the inherent negativity of the initial prediction. Given that, the examples in Figure 6 show that in most cases the sum of the features pushes toward a class, while few Mordred descriptors and receptor associations could remarkably alter the result of the prediction.

Regarding the implementation of the Graph Convolutional Neural Network (GCN), we converted the dataset of molecules into a graph-based representation using NetworkX. Each node and edge of the graphs are described by a specific set of features as listed in Table S1. The developed GCN model achieved on the test set ROC AUC and PR AUC scores of 0.88 and 0.67, respectively (Figure 7). In terms of AUC, the TML outperformed the GCN model reaching 0.92 and 0.75 for ROC AUC and PR AUC scores, respectively (Table 1). It is worth noticing that TML and GCN have comparable performance in the negative samples (class 0), whereas the precision of the TML in the positive samples (class 1) is remarkably higher (TML: 0.78; GCN: 0.62). This discrepancy in performance could be attributed to the dataset's imbalance, where there are nearly five times more negative pairs compared to positive samples. This imbalance likely affects the performance, resulting in lower precision for the positive class compared to the negative class. This issue is exacerbated for the GCN model due to its increased sensitivity to limited samples in the training set.

The GNNExplainer (Ying et al., 2019) and Grad-CAM (Selvaraju et al., 2020) methods were employed to explain the GCN model. GNNExplainer offers insights at a graph level, considering the entirety of the molecule, while Grad-CAM can highlight the node importance and provide information on its impact towards positive or negative class. The GNNExplainer on a graph level can easily compared with the Tree-based feature importances. Noticeably, features of the molecule directly linked to specific types of interactions (electrostatics, polar interactions, hydrogen bonds, etc.) with the receptor are identified as the most important by the GNNExplainer. In particular, for the strychnine molecule (Figure 8A), the most important node features include the atom partial charge (Gasteiger Charge) and partition coefficient (logP), which can be related to the hydrophilicity of the molecule, a crucial factor for the ligand-target interaction. Moreover, edge importance emphasises the structural motifs that affect the model's output (Figure 8B). On the contrary, Grad-CAM and UGrad-CAM can generate heatmaps of gradient-weighted class activations, offering a robust visual explanation of the atoms influencing the prediction and indicating which class each node is favouring. Figure 8C shows UGrad-CAM applied on the Strychnine-TAS2R46 pair. This association was chosen specifically because of recent experimental evidence regarding strychnine's binding to TAS2R46 (Xu et al., 2022). The authors suggest that the binding occurs thanks to a π-π interaction between residue W88 and strychnine's benzene ring, and with the formation hydrogen bond between residue E265 and one of the two tertiary amines of strychnine. By looking at the UGrad-CAM plot (Figure 8C), it is interesting to notice that the same tertiary amine involved in the binding is one of the most activated atoms towards class 1 prediction. Interestingly, the GNNExplainer analysis also consistently identified edges connected to this tertiary amine as crucial for the final prediction (Figure 8B). In summary, the explainability of the GCN model appears particularly promising in identifying the structural molecular features underlying the ligand-receptor association and facilitating the explanation of the model.

The proposed models have been also compared with previous examples in the literature. At present, as far as the authors know, only three methods, i.e. BitterX (Huang et al., 2016), BitterSweet (Tuwani et al., 2019), and BitterMatch (Margulis et al., 2022), have been developed to predict the interaction between bitterants and TAS2Rs. We exclusively compared our model with BitterMatch as it is the most recently developed tool and has demonstrated the best performance. Additionally, while the BitterSweet web server provides predictions on bitterant-TAS2R associations, no information is available regarding the specific development of the BitterSweet model for these associations in either the original publication or the online materials. Moreover, BitterX was not considered for a fair comparison because it was trained on a profoundly different dataset (540 bitterants with 260 positive and 260 negative bitterant-TAS2R interactions). To effectively compare BitterMatch with the present work, the BitterMatch model from the official GitHub repository (https://github.com/YuliSl/BitterMatch) was trained only on human data (*BM Human-Only*) by removing data relative to murine receptors. The comparison of the AUC and PR curves for the TML, GCN and BM-Human-Only models are reported in Figure S5, whereas all performance metrics are summarised in Table S3. As indicated in the comparison between the TML and GCN models, performance metrics for class 0 are higher and almost identical, while performance for class 1 is notably lower and with some differences for the three compared models. Consistent with our observations in the comparison between the TML and GCN models, the performance on class 0 (negative associations) is again notably higher than on class 1 (positive associations) for the BitterMatch (BM) model. This disparity can again be attributed to the dataset being unbalanced towards the negative class. Overall, the three models achieved similar PR AUC scores on their test sets, and the other metrics indicated that the three models are nearly equivalent, with BitterMatch achieving slightly less performance in Recall, F1, and F2 in class 1. However, it is worth mentioning that, unlike BitterMatch, the present models are capable of predicting the class of any query molecule within the model's applicability domain using only its SMILES representation. This expands their usage to a wider audience and paves the way toward the possibility of using these methodologies in emerging fields of interest, such as precision nutrition or nutraceutical development.

In summary, when considering only performances on prediction, the TML model achieves a better Precision score when compared to the GCN model, also offering a better alternative to other state-of-the-art models like BitterMatch in terms of Recall, F1 and F2 scores. On the other hand, GCN possesses the ability to generate more impactful visual explanations, which could be easily interpreted and exploited for subsequent tasks. Hence, we regard the Traditional Machine Learning (TML) and Graph Convolutional Network (GCN) models presented in this study as valuable complementary tools, offering robust predictions of TAS2Rs-bitterants interactions along with insightful visual explanations that connect molecular structures with their targeting capabilities.

## Conclusion

This work presents a novel approach to predicting the interactions between bitter taste receptors and their ligands, using both a Traditional Machine-Learning (TML) approach and Graph Convolution Neural Networks (GCN). Both models were also designed to be explainable, either by using interpretable algorithms, such as decision trees, or by applying custom methods, such as SHAP, GNNExplainer, and Grad-CAM, to reveal the most important features and molecular motifs underlying the predictions. The results indicated that both the TML and GCN models achieved similar and satisfactory predictive performance. Compared to previous works, the developed models offer ease of use, applicability to new molecules within the model's applicability domain, and interpretability, providing valuable insights into the associations between molecular features and their corresponding bitter taste receptor targets. While the TML model demonstrated slightly superior reliability in terms of performance metrics, the GCN model offers explanations through visually rich representations directly of the molecular structure of the compounds under investigation, thereby enhancing the interpretability of model predictions.

In a broader context, the *in silico* identification of promising compounds capable of targeting TAS2Rs holds significant potential across multiple fields. These methodologies can aid in the development of novel bitter modulators and enhancers, exerting a substantial impact not only in the food industry but also in the pharmaceutical sector, where the bitter taste of medications often poses challenges to effective administration for children or the elderly. Furthermore, since certain bitter receptors are also found in extra-oral tissues and are implicated in various diseases such as obesity, diabetes, asthma, and cancer, the development of these machine-learning models represents a promising strategy for identifying specific ligands capable of targeting these proteins. Such models have the potential to overcome some of the limitations associated with traditional experimental approaches, including high cost, low throughput, and ethical issues.

# Acknowledgements

The present work has been developed as part of the VIRTUOUS project, funded by the European Union's Horizon 2020 research and innovation program under the Marie Sklodowska-Curie-RISE Grant Agreement No. 872181 (https://www.virtuoush2020.com/).

The authors also acknowledge the CINECA award under the ISCRA initiative for the availability of high-performance computing resources and support.

The authors would like to acknowledge Marcello Miceli (Politecnico di Torino, Turin, Italy) for valuable feedback and helpful discussions during the definition and realisation of the work.

# Author Contributions

**Francesco Ferri**: Conceptualization; Data curation; Formal analysis; Methodology; Software; Validation; Visualization; Writing - original draft; Writing - review & editing.

**Marco Cannariato**: Conceptualization; Data curation; Formal analysis; Methodology; Software; Validation; Visualization; Writing - original draft; Writing - review & editing.

**Lorenzo Pallante**: Conceptualization; Data curation; Visualization; Writing - original draft; Writing - review & editing.

**Eric A. Zizzi**: Conceptualization; Supervision; Writing - review & editing;

**Marcello Miceli**: Conceptualization; Supervision; Writing - review & editing;

**Giacomo Di Benedetto**: Supervision; Writing - review & editing;

**Marco Agostino Deriu**: Conceptualization; Funding acquisition; Project administration; Resources; Supervision; Validation; Writing - review & editing.

# Data Availability

All codes and data needed to replicate the results of the present work are accessible at https://github.com/francescofers/TAS2RsPredictor.

# Competing Interests Statement

The authors declare no competing interests.

predictor using a multi-objective machine learning approach. *Scientific Reports*, *12*(1), 21735. https://doi.org/10.1038/s41598-022-25935-3

Pallante, L., Malavolta, M., Grasso, G., Korfiati, A., Mavroudi, S., Mavkov, B., Kalogeras, A., Alexakos, C., Martos, V., Amoroso, D., di Benedetto, G., Piga, D., Theofilatos, K., & Deriu, M. A. (2021). On the human taste perception: Molecular-level understanding empowered by computational methods. *Trends in Food Science & Technology*, *116*, 445–459. https://doi.org/10.1016/j.tifs.2021.07.013

Paszke, A., Gross, S., Massa, F., Lerer, A., Bradbury, J., Chanan, G., Killeen, T., Lin, Z., Gimelshein, N., Antiga, L., Desmaison, A., Köpf, A., Yang, E., DeVito, Z., Raison, M., Tejani, A., Chilamkurthy, S., Steiner, B., Fang, L., … Chintala, S. (2019). *PyTorch: An Imperative Style, High-Performance Deep Learning Library* (arXiv:1912.01703). arXiv. http://arxiv.org/abs/1912.01703

Patro, S. G. K., & Sahu, K. K. (2015). *Normalization: A Preprocessing Stage* (arXiv:1503.06462). arXiv. https://doi.org/10.48550/arXiv.1503.06462

Pope, P. E., Kolouri, S., Rostami, M., Martin, C. E., & Hoffmann, H. (2019). Explainability Methods for Graph Convolutional Neural Networks. *2019 IEEE/CVF Conference on Computer Vision and Pattern Recognition (CVPR)*, 10764–10773. https://doi.org/10.1109/CVPR.2019.01103

Risso, D., Drayna, D., & Morini, G. (2020). Alteration, Reduction and Taste Loss: Main Causes and Potential Implications on Dietary Habits. *Nutrients*, *12*(11), Article 11. https://doi.org/10.3390/nu12113284

Rogers, D. J., & Tanimoto, T. T. (1960). A Computer Program for Classifying Plants. *Science*, *132*(3434), 1115–1118. https://doi.org/10.1126/science.132.3434.1115

Selvaraju, R. R., Cogswell, M., Das, A., Vedantam, R., Parikh, D., & Batra, D. (2020). Grad-CAM: Visual Explanations from Deep Networks via Gradient-Based Localization. *International*
24

# Explainable Machine Learning and Deep Learning Models for Predicting TAS2R-Bitter Molecule Interactions

Francesco Ferri[1], Marco Cannariato[1], Lorenzo Pallante[1], Eric A. Zizzi[1], Marcello Miceli[1], Giacomo di Benedetto[2] and Marco A. Deriu[1*]

[1] Politecnico di Torino, PolitoBIOMedLab, Department of Mechanical and Aerospace Engineering, Torino, 10129, Italy
[2] 7hc srl, Rome, Italy

* marco.deriu@polito.it

# Supporting Information

## Material and methods

### Min-Max Normalisation

All non-binary data was normalized with Min-Max normalization, returning values between 0 and 1 [1]. Let $A'$ contain all the Min-Max Normalized data, $A$ be the original data, $C = 0$ and $D = 0$ be the pre-defined boundary of the normalized data, Min-Max normalization is defined as:

$$A' = \frac{A - minimum\ value\ of\ A}{maximum\ value\ of\ A - minimum\ value\ of\ A} * (D - C) + C$$

### GCN model

#### Node and Edge features

Table S1 summarises node and edge features used for graph representation in the GCN model. Most common features, such as atom weight and bond types, and other optional features are included, in line with previous literature [2], [3], [4].

*Table S1. List of all node (atom) and edge (bond) features employed in the GNN; \* if normalized with Min-Max normalization [0,1]; ° indicates a Boolean feature (0 or 1); ^ if one-hot encoded.*

| #  | Node features                                                              | Edge features   |
|----|----------------------------------------------------------------------------|-----------------|
| 1  | Mass* = normalized mass (on Iodium mass)                                   | Single bond°    |
| 2  | logP* = atom contribution to logP of the molecule                          | Double bond°    |
| 3  | MR* = atom contribution to Molar Refractivity of the molecule              | Triple bond°    |
| 4  | Estate* = atom contribution to EState of the molecule                      | Aromatic bond°  |
| 5  | ASA* = atom contribution to the Accessible Solvent Area of the molecule    |                 |
| 6  | TPSA* = atom contribution to the Topological Polar Surface Area of the molecule |             |
| 7  | Partial Charge* = Atom partial charge                                      |                 |
| 8  | Degree^ = number of directly bonded neighbours to the atom                 |                 |
| 9  | Implicit Valence^ = number of implicit hydrogens on the atom               |                 |
| 10 | nH^ = number of total hydrogens on the atom                                |                 |

| 11    | Aromatic° = if the atom is part of an aromatic ring |
|-------|-----------------------------------------------------|
| 12-34 | Receptor Association^                               |

## Evaluation metrics

Let True Positive (TP) be the number of positive cases correctly identified as positives; False Negative (FN) be the number of positive cases that are erroneously classified as negative cases; False Positive (FP) be the number of negative cases misclassified as positive cases; True Negative (TN) be the number of negative cases rightfully classified as negative cases. The basic metrics used for the evaluation of a model are:

$$Precision = \frac{TP}{TP + FP}$$

$$Recall\ (sensitivity) = \frac{TP}{TP + FN}$$

$$Specificity = \frac{TN}{TN + FP}$$

A perfect model will identify all positive examples ($Recall = 1$), and score only the truly positive examples ($Precision = 1$).

$$F_\beta = \frac{(1 + \beta^2) \times Recall \times Precision}{(\beta^2 \times Precision) + Recall}$$

β-varied F-measure uses a coefficient β to balance the relative importance of precision and recall. A lower β gives less weight to precision, while a higher β gives more weight to it.

Let $P_n$ and $R_n$ be the precision and the recall at the $n$-th threshold, respectively, Average Precision is then defined as:

$$Average\ Precision = \sum_n (R_n - R_{n-1}) P_n$$

Moreover, a common way to assess binary decision problems is to use Receiver Operating Characteristic (ROC) curves, as they show how true positives and false positives vary. However, ROC curves are not reliable when the classes are unbalanced, because they ignore how false positives affect the overall performance. Precision-Recall (PR) curves are better for imbalanced class distributions because they capture the precision and recall of the algorithm. The Area Under the Curve (AUC) is a numerical indicator of the ROC and PR curve trend, that can summarize the performance of a classifier into a single number.

# Results

## TML approach

**Comparison of traditional machine-learning methods**

Several traditional machine-learning-based algorithms, such as Gaussian Naïve Bayer, Logistic Regression, KNeighbors, SVM, and RandomForest classifiers, were compared to find the best model. The resulting ROC curves on the validation set (10-fold CV) are reported in Figure S1.

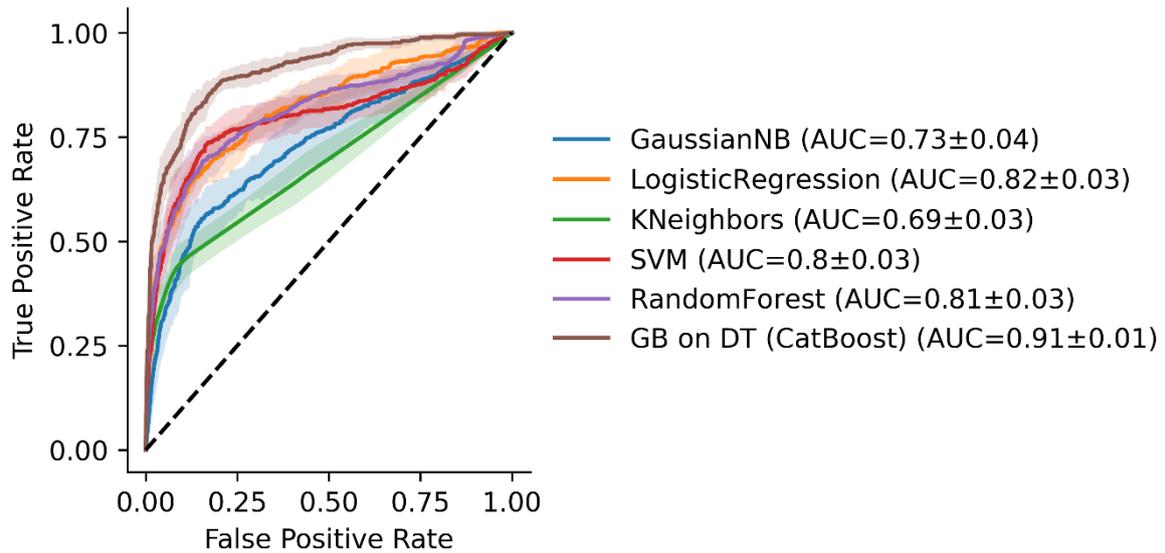

*Figure S1. ROC curves on Validation set for different classical ML algorithms. The mean values for the 10-fold cross-validation are depicted as solid lines, while the standard deviations are represented by shaded areas.*

**CatBoostClassifier hyper-parameters**

The tuned hyper-parameters for the TML approach are summarised in Table S2.

*Table S2. CatBoostClassifier hyper-parameters.*

| Boosting Type | Depth | Iterations | Learning Rate | Leaf Estimation Iterations | L2 Leaf Reg | Subsample |
|---|---|---|---|---|---|---|
| Plain | 6 | 1000 | 0.1 | 4 | 3 | 0.7 |

## GCN model
### GB on DTs – Backward Sequential Feature Selection (SFS) method

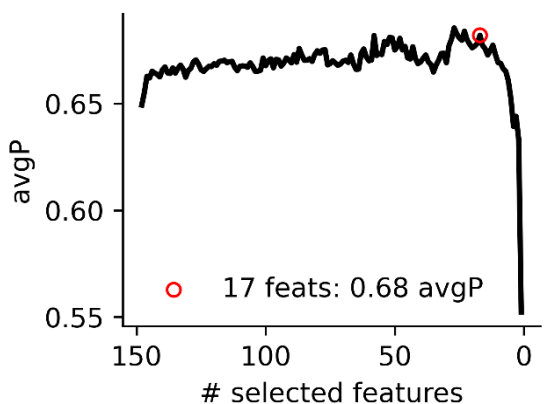

*Figure S2. Number of features selected vs average precision value during Backward-SFS process. The best average precision value was achieved using 17 features.*

### Comparison of the feature selection methods

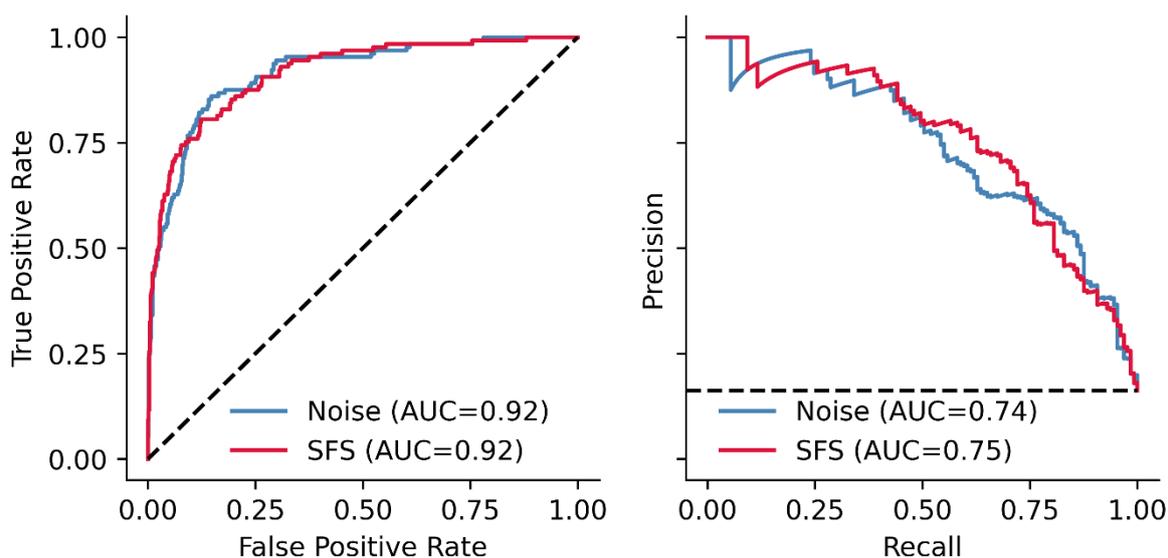

*Figure S3. Performance comparison between the "noisy" feature selection and the Backward-SFS methods.*

# Discussion

## Applicability Domain (AD)

The applicability domain (AD) was implemented similarly to previous literature [5], [6] and our previous works [7], [8]. Specifically, the applicability domain (AD) was delineated through an average similarity approach. The underlying concept is that compounds exhibiting high dissimilarity to those present in the training set should be categorized as outside the AD of the developed models. The average similarity score between test and training compounds is calculated as follows: (i) the Morgan Fingerprints (1024 bits, radius 2) were calculated using RDKit for all the compounds in the training set; (ii) a similarity score was then evaluated between each molecule in the training and test sets and the previously-defined fingerprints using the Jaccard similarity index from RDKit; (iii) then the average similarity score was computed by averaging the similarity scores of the 5 most similar couple of compounds. The distribution of the average similarity scores for the training and test sets was used to identify a similarity threshold to discriminate between query compounds inside or outside the domain of applicability of the developed model (Figure S4).

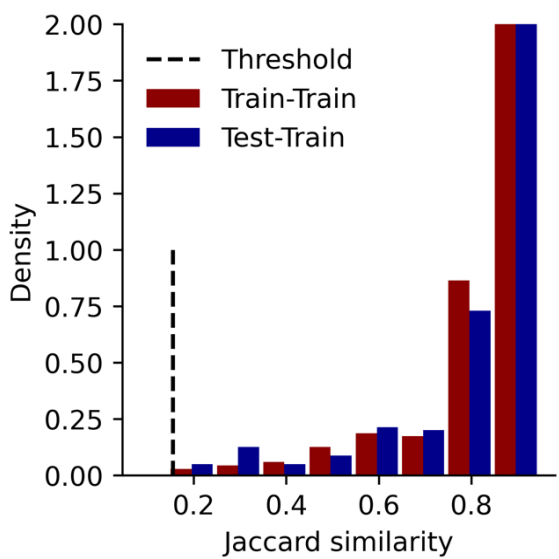

*Figure S4. Histograms of average similarity scores of training and test sets. The average similarity score is derived by averaging the Jaccard similarity score between the five most similar compounds in the training set. The red histograms represent the distribution of the average similarity scores for all the compounds composing the training set, whereas the blue histogram is the distribution for the test set. The lower limit of the above-mentioned distributions allows for determining the similarity threshold of the applicability domain.*

## Models' comparison with BitterMatch

To compare the developed models with existing literature, the BitterMatch model was considered. However, to perform a better comparison, as the BitterMatch model was trained also on murine data, we trained the architecture proposed by Margulis et al. after having removed from their initial dataset all columns and ligands relative to murine receptors. The developed model, which has been named HumanOnly-BitterMatch (BM), was obtained following BitterMatch's official GitHub code [9]. The *new-ligands* scenario was selected since it is more similar to the presented approaches. Notably, we inserted the same random seed used in the rest of the work to achieve better reproducibility of the study, as no random seed for the train-test split was originally present in the official code. BM Human-Only performances were compared against the developed models, considering the average performances on 100 bootstrapped versions of the test set (resampling of 90%). The obtained PR curves are shown in Figure S5, while the details of the models' performances are reported in Table S3. To be noted, the two datasets differ, as TML is trained on an expanded dataset.

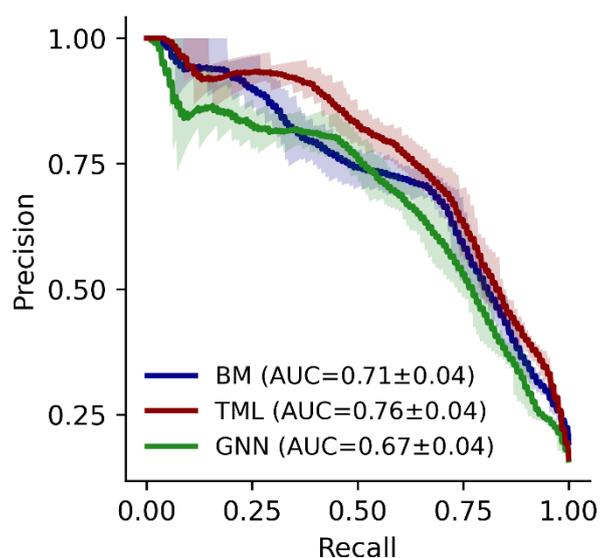

*Figure S5. Comparison between the PR curves of the BitterMatch model trained on human-only data (BM, blue), the TML model (red), and the GCN model (green), obtained through a bootstrap of the test set. The mean PR curves are represented as a continuous line, while the shaded parts correspond to the region between the 25th and 75th percentiles.*

Table S3. Comparison of the performance of the TML, GCN, and human-only BitterMatch (BM) models on the test set. Values in bold represent the highest among the metrics compared.

|  |  | TML | GCN | BM |
|---|---|---|---|---|
| Class 0 | Precision | 0.93 | **0.94** | 0.88 |
|  | Recall | **0.97** | 0.92 | 0.96 |
|  | F1 | **0.95** | 0.93 | 0.92 |
|  | F2 | **0.96** | 0.93 | 0.95 |
| Class 1 | Precision | **0.78** | 0.62 | 0.75 |
|  | Recall | 0.60 | **0.67** | 0.44 |
|  | F1 | **0.68** | 0.64 | 0.55 |
|  | F2 | 0.63 | **0.66** | 0.48 |